\documentstyle[11pt,amssymb]{article}

\makeatletter
\@addtoreset{equation}{section}
\makeatother

\def\ben{\begin{equation}}
\def\een{\end{equation}}

\let\a=\alpha    \let\e=\epsilon
    
  \let\n=\nu

\let\C=\Chi

\def\nn{\nonumber} \def\bd{\begin{document}} \def\ed{\end{document}}
\def\ds{\documentstyle} \let\fr=\frac \let\bl=\bigl \let\br=\bigr
\let\Br=\Bigr \let\Bl=\Bigl
\let\bm=\bibitem
\let\na=\nabla
\let\pa=\partial \let\ov=\overline
\newcommand{\be}{\begin{equation}}
\newcommand{\ee}{\end{equation}}
\def\ba{\begin{array}}
\def\ea{\end{array}}
\def\ft#1#2{{\textstyle{{\scriptstyle #1}\over {\scriptstyle #2}}}}
\def\fft#1#2{{#1 \over #2}}
\def\del{\partial}
\def\vp{\varphi}
\def\sst#1{{\scriptscriptstyle #1}}
\def\oneone{\rlap 1\mkern4mu{\rm l}}
\def\td{\tilde}
\def\wtd{\widetilde}
\def\ie{\rm i.e.\ }
\def\dalemb#1#2{{\vbox{\hrule height .#2pt
        \hbox{\vrule width.#2pt height#1pt \kern#1pt
                \vrule width.#2pt}
        \hrule height.#2pt}}}
\def\square{\mathord{\dalemb{6.8}{7}\hbox{\hskip1pt}}}
\newcommand{\ho}[1]{$\, ^{#1}$}
\newcommand{\hoch}[1]{$\, ^{#1}$}
\newcommand{\bea}{\begin{eqnarray}}
\newcommand{\eea}{\end{eqnarray}}
\newcommand{\ra}{\rightarrow}
\newcommand{\lra}{\longrightarrow}
\newcommand{\Lra}{\Leftrightarrow}
\newcommand{\ap}{\alpha^\prime}
\newcommand{\bp}{\tilde \beta^\prime}
\newcommand{\tr}{{\rm tr} }
\newcommand{\Tr}{{\rm Tr} }
\def\0{{\sst{(0)}}}
\def\1{{\sst{(1)}}}
\def\2{{\sst{(2)}}}
\def\3{{\sst{(3)}}}
\def\4{{\sst{(4)}}}
\def\5{{\sst{(5)}}}
\def\6{{\sst{(6)}}}
\def\7{{\sst{(7)}}}
\def\8{{\sst{(8)}}}
\def\n{{\sst{(n)}}}
\def\cA{{{\cal A}}}
\def\cF{{{\cal F}}}
\def\tV{\widetilde V}
\def\tW{\widetilde W}
\def\tH{\widetilde H}
\def\tE{\widetilde E}
\def\tF{\widetilde F}
\def\tA{\widetilde A}
\def\im{{{\rm i}}}
\def\tY{{{\wtd Y}}}
\def\ep{{\epsilon}}
\def\vep{{\varepsilon}}
\def\R{\rlap{\rm I}\mkern3mu{\rm R}}
\def\bD{{{\bar D}}}
\def\alp{{{\a'}^3}}
\def\R{\rlap{\rm I}\mkern3mu{\rm R}}
\def\bD{{{\bar D}}}
\def\R{{{\Bbb R}}}
\def\C{{{\Bbb C}}}
\def\H{{{\Bbb H}}}
\def\CP{{{\Bbb C}{\Bbb P}}}
\def\RP{{{\Bbb R}{\Bbb P}}}
\def\Z{{{\Bbb Z}}}
\def\bA{{{\Bbb A}}}
\def\bB{{{\Bbb B}}}
\def\bC{{{\Bbb C}}}
\def\bR{{{\Bbb R}}}
\def\bD{{{\Bbb D}}}
\def\bE{{{\Bbb E}}}
\def\bZ{{{\Bbb Z}}}
\def\Re{{{\frak{Re}}}}
\def\Im{{{\frak{Im}}}}
\def\cosec{{\,\hbox{cosec}\,}}
\def\Gm{{\Gamma_{\!\! -}}}
\def\Gp{{\Gamma_{\!\! +}}}
\def\stan{{standard }}
\def\nonstan{{supernumerary }}

\def\cosech{{\hbox{cosech}}}

\def\etcyc{{\hbox{and cyclic}}}
\def\btheta{{\bar\theta}}

\newcommand{\tamphys}{\it Center for Theoretical Physics,
Texas A\&M University, College Station, TX 77843, USA}

\newcommand{\mitchell}{\it George P. \& Cynthia W.
Mitchell Institute for Fundamental Physics,\\
Texas A\&M University, College Station, TX 77843-4242, USA}
\newcommand{\umich}{\it Michigan Center for Theoretical Physics,
University of Michigan\\ Ann Arbor, MI 48109, USA}
\newcommand{\upenn}{\it Department of Physics and Astronomy,
University of Pennsylvania, Philadelphia,  PA 19104, USA}
\newcommand{\SISSA}{\it  SISSA-ISAS and INFN, Sezione di Trieste\\
Via Beirut 2-4, I-34013, Trieste, Italy}

\newcommand{\newton}{\it Isaac Newton Institute for Mathematical Sciences,\\
20 Clarkson Road,  University of Cambridge,
Cambridge CB3 0EH, UK}

\newcommand{\ihp}{\it Institut Henri Poincar\'e\\
  11 rue Pierre et Marie Curie, F 75231 Paris Cedex 05}

\newcommand{\damtp}{\it DAMTP, Centre for Mathematical Sciences,
 Cambridge University\\  Wilberforce Road, Cambridge CB3 OWA, UK}
\newcommand{\itp}{\it Institute for Theoretical Physics, University of
California\\ Santa Barbara, CA 93106, USA}

\newcommand{\imperial}{\it The Blackett Laboratory, Imperial College London\\
Prince Consort Road, London SW7 2AZ. }

\newcommand{\auth}{
H. L\"u\hoch{\ddagger1},
C.N. Pope\hoch{\ddagger1} and K.S. Stelle\hoch{\star2} }

\begin{document}
\begin{flushright}
\hfill{
MIFP-06-28 }\\
\hfill{
\bf hep-th/0611299}
\end{flushright} 

\begin{center}  

{\Large {\bf Consistent Pauli Sphere Reductions and the Action 
}}   

\vspace{15pt}

\auth

\vspace{7pt}
{\hoch{\ddagger}\mitchell}

\vspace{7pt}
{\hoch{\star}\imperial} 

\vspace{30pt}

\underline{ABSTRACT}
\end{center}  

   It is a commonly held belief that a consistent dimensional reduction ansatz
can be equally well substituted into either the higher-dimensional 
equations of motion or the higher-dimensional action, and that the 
resulting lower-dimensional theories will be the same.  This is certainly
true for Kaluza-Klein circle reductions and for DeWitt group-manifold
reductions, where group-invariance arguments guarantee the equivalence.
In this paper we address the question in the case of the non-trivial
consistent Pauli coset reductions, such as the $S^7$ and $S^4$ reductions
of eleven-dimensional supergravity.  These always work at the level of the
equations of motion.  In some cases the reduction ansatz 
can only be given at the level of field strengths, rather than the
gauge potentials which are the fundamental fields in the action, and so
in such cases there is certainly no question of being able to substitute 
instead into the action.  By examining explicit examples, we show that
even in cases where the ansatz can be given for the fundamental fields
appearing in an action, substituting it into the higher-dimensional 
action may not give the correct lower-dimensional theory.  This highlights
the fact that much remains to be understood about the way in which 
Pauli reductions work.

{\vfill\leftline{}\vfill
\vskip 10pt \footnoterule
{\footnotesize \hoch{1}
Research supported in part by DOE grant DE-FG03-95ER40917
\vskip -12pt} \vskip 14pt
{\footnotesize \hoch{2}
Research supported in part by the EC under TMR
contract HPRN-CT-2000-00131 and by PPARC \\
$\phantom{xxxx}$ under SPG grant
PPA/G/S/1998/00613.
\vskip -12pt}  \vskip  14pt
}

\pagebreak
\setcounter{page}{1}

\tableofcontents
\addtocontents{toc}{\protect\setcounter{tocdepth}{2}}

\section{Introduction}

   The idea of dimensional reduction in general relativity
was pioneered by Kaluza in 1921 \cite{kaluza}.  He observed
that by taking the metric in five dimensions to be
independent of the fifth coordinate, one obtained a four-dimensional
system describing electromagnetism with an additional scalar field,
coupled to four-dimensional gravity. It was later observed by Klein
that one could take the fifth coordinate to be compactified into a
circle, which would become effectively unobservable if its radius were
taken to be sufficiently small \cite{klein}.\footnote{However, at the same
time Klein took the retrograde step of setting the scalar
field to a constant, in an attempt to obtain pure Einstein-Maxwell 
theory in four dimensions.  (See, for example, \cite{rafneu}.)  
The problem with doing this is that 
it contradicts the scalar field's own equation of motion, which has  
a source term built from the square of the Maxwell field strength.  
This is the {\it ur} example of an inconsistent
reduction ansatz, although the remedy in this case is simple; do not
set the scalar to a constant.}  By a straightforward
process of iteration, one can immediately extend the idea of Kaluza-Klein
reduction to the case of compactification of a higher-dimensional
theory on the $n$-dimensional torus, $T^n$.  The lower-dimensional 
theory will include the gauge fields of $U(1)^n$, associated with the
$U(1)^n$ isometry group of $T^n$.

   The next significant development in the idea of obtaining gauge fields
from the isometry group of a compact internal space was the proposal
by Pauli, in 1953, that one might obtain the non-abelian gauge fields
of the group $SO(3)$ by compactifying on the 2-sphere \cite{pauli}.  He
had in mind the idea of reducing six-dimensional general relativity
to four dimensions.
More generally, one could envisage obtaining the Yang-Mills gauge group
$G$ by means of a compactification on the coset space $G/H$.  Following
the discussion in \cite{cvgilupo}, we shall refer to any such reduction
as a {\it Pauli Reduction}.  Although Pauli proposed reductions of this
kind, he also recognised that there would be difficulties in realising 
them in practice.  These difficulties are associated with
what is nowadays called an ``inconsistency of the reduction ansatz.''
For example, if one attempts to implement Pauli's original idea of
reducing six-dimensional gravity on $S^2$, one finds that one cannot
obtain consistent four-dimensional equations of motion by substituting
the ansatz into the six-dimensional Einstein equations.

   The definition of a ``consistent reduction ansatz'' is one that can
be substituted into the higher-dimensional equations of motion, yielding
a consistent system of equations for the lower-dimensional system, with
the consequent property that any solution of the lower-dimensional 
equations gives rise to a solution of the original higher-dimensional
equations.  The obstacle to achieving such a consistent reduction, in
a situation of the type envisaged by Pauli, is that one will not
in general be able to ``factor out'' the dependence on the coordinates
of the internal reduction space when one substitutes into the 
higher-dimensional equations of motion.  

   Another way of characterising the inconsistency is as follows.  One
could always choose to perform an expansion of all the higher-dimensional
fields in terms of complete sets of appropriate harmonic fields on the
internal space.  This would just give a generalised Fourier expansion of
the original theory, resulting in a theory containing
infinite sets of lower-dimensional fields.  This would necessarily be
consistent, in the sense defined above.  The lower-dimensional fields
would essentially comprise a finite set of massless fields, coupled
to infinite towers of massive fields, with masses unbounded above.  In the
type of reduction envisaged by Pauli, the reduction ansatz would only
retain the massless fields (or, at least, some finite subset, including
the metric and the Yang-Mills gauge fields of the isometry group of
the internal space).  The crucial question, therefore, is whether setting
the remainder of the infinite towers of fields to zero is consistent
with their own equations of motion.  The answer, in the case of a ``Pauli
reduction'' of a generic higher-dimensional theory, is an unequivocal
``No.''  One cannot, in general, consistently set the infinite towers to
zero, and so the Pauli reduction will in general be 
inconsistent.\footnote{Note that we are concerned here with exact 
mathematical statements.  We are not interested here in the question of 
whether
the massive towers can be neglected, or ``integrated out,'' in some
physically-motivated low-energy approximation.}

   The inconsistency of a generic Pauli reduction can be attributed to
the fact that the internal-space harmonics associated with the infinite
massive towers are generated by products of the internal-space
harmonics of the retained ``massless'' sector.  This implies, in view
of the non-linear nature of the theory one is reducing, that there will
be source terms in the equations of motion for the massive towers,
comprised of products of the purely massless fields.  Thus the massive towers
cannot be consistently set to zero.  If one starts at the outset with an
ansatz involving only the massless sector, the inconsistency will reveal
itself as an inability to extract purely lower-dimensional equations of
motion with the coordinate dependence of the internal space factored out.
Instead, one obtains an over-constrained system of equations with
no sensible lower-dimensional content.

   The inconsistencies described above are of course avoided in the
Kaluza-Klein reductions on $S^1$ or $T^n$.  An easy way to see this
is that the massive fields in the Kaluza-Klein towers are all {\it charged}
under the $U(1)^n$ gauge group, whilst the massless modes are
uncharged (\ie they are independent of the toroidal coordinates).  Since
products of the uncharged massless fields can never carry charge, it is
impossible for them to act as sources for the massive charged fields.

   Another situation where one is guaranteed a consistent reduction is
in the scheme introduced by DeWitt in 1963, in which a reduction is
performed on the compact group manifold $G$ \cite{dewitt}.  The bi-invariant 
metric on $G$ has the isometry group $G_L\times G_R$, denoting the
independent left and right actions of $G$ on the group manifold.  Thus
in a generalised Fourier expansion one would obtain massless Yang-Mills
gauge fields of the full $G_L\times G_R$ isometry group in the lower
dimension.  In the DeWitt reduction, however, only the gauge fields of
one copy, say $G_L$, are retained.  In fact, the full set of fields 
retained in the DeWitt reduction comprise precisely {\it all} those
which are {\it singlets} under $G_R$.  Since $G_R$ acts transitively on
$G$, this means that the retained fields will be finite in number.  The
DeWitt reduction will necessarily be a consistent one, for a group
theoretic reason analogous to the one discussed above for Kaluza-Klein $T^n$
reductions.  Namely, since all the retained fields are singlets under
$G_R$, it is impossible for products of these fields to act as sources for
the massive towers that are set to zero, since by definition all the
fields in those towers are non-singlets under $G_R$.

    It is almost a truism in any of the consistent reductions that has
such a group-theoretic explanation that one could equally well choose
to substitute the reduction ansatz into the higher-dimensional action,
integrate over the internal group manifold, and thereby obtain a 
lower-dimensional action whose equations of motion coincide with those
one would obtain by instead substituting the ansatz into the 
higher-dimensional equations of motion.  It is presumably for this 
reason that a ``folklore'' belief has arisen that in any consistent
reduction, whether or not the consistency has a group-theoretic
explanation, one should be able equivalently to substitute the ansatz
into either the higher-dimensional equations of motion or instead into
the higher-dimensional action.  

    In fact in the case of consistent Pauli
reductions, where no group-theoretic explanation for the consistency
exists, it is not in general true that one obtains the same 
lower-dimensional theory by substituting the reduction ansatz into 
either the equations of motion or the action of the higher-dimensional
theory.  Indeed, in many cases the basic premise underlying the question
is false; one may well not even be able to write the ansatz in a form that
would allow it to be substituted into the action, and so the question of
the equivalence or otherwise of the two procedures does not even arise.
Such a situation typically occurs if the ansatz can be expressed only at the
level of the field strengths of the higher-dimensional theory, but not
in terms of the underlying gauge potentials that are the fundamental 
fields in the action.  Even if one looks at examples where the ansatz
can be given for the gauge potentials themselves, it can still be the
case that substitution into the higher-dimensional action fails to give
the correct lower-dimensional action.\footnote{An unambiguous definition
of the ``correct'' lower-dimensional equations is the ones such that,
for any solution of these equations, the reduction ansatz yields a 
solution of the higher-dimensional equations of motion.}

    The purpose of the present paper is to explore in detailed examples
this question of
the equivalence, or otherwise, of substituting a consistent Pauli reduction
ansatz into the higher-dimensional equations of motion or action.  Examples
of consistent Pauli reductions are very few and far between, and in fact
most of the known cases are associated with supergravity theories.  The
reason for the paucity of examples can be understood using an argument
given in \cite{clpst}.  If the consistent Pauli reduction of a particular 
theory on $S^n$
is possible, then by definition it will yield a lower-dimensional theory 
with Yang-Mills fields gauging the group $SO(n+1)$.  By turning off the
gauge coupling, which amounts to sending the radius of the sphere to
infinity, the lower-dimensional theory will become the one that would
instead result from a Kaluza-Klein reduction on $T^n$.  From this 
standpoint, the $S^n$-reduced theory can then be viewed as a gauging of
an $SO(n+1)$ global symmetry of the $T^n$-reduced theory.  A {\it
necessary} condition for the consistency of the $S^n$ Pauli reduction is
therefore that if the higher-dimensional theory is instead reduced on
$T^n$, it must give rise to a lower-dimensional theory with a global
symmetry group $G$ that is {\it at least} large enough to contain
$SO(n+1)$.   Now, a generic theory reduced on $T^n$ will have only
a $GL(n,\R)$ global symmetry, and this has as maximal compact subgroup
$SO(n)$, which of course does not contain $SO(n+1)$.  The upshot is that
only in the case of a higher-dimensional theory whose $T^n$ reduction
gives a theory with a {\it global symmetry enhancement}, to a group $G$
that contains $SO(n+1)$, can there be any chance that a Pauli reduction
on $S^n$ is possible.  In practice, most of the examples where such
a global symmetry enhancement occurs are found in supergravity, such
as the $T^n$ reductions of $D=11$ supergravity, which yield maximal 
supergravities in dimension $D=11-n$ with $E_n$ global symmetry.     

   Three important examples of Pauli reductions are the $S^4$ and
$S^7$ reductions of eleven-dimensional supergravity, and the $S^5$
reduction of type IIB supergravity.  Each of these satisfies the necessary
condition discussed above; the $S^n$ isometry groups, $T^n$-reduced
global symmetry groups and  maximal compact subgroups are as follows:
\bea
S^4:&& SO(5) \subset SO(5) \subset SL(5,\R)\,,\nn\\
S^7:&& SO(8) \subset SU(8) \subset E_7\,,\nn\\
S^5:&& SO(6)\subset USp(8) \subset E_6\,.
\eea
The consistency of the $S^4$ Pauli reduction was demonstrated in 
\cite{nasvamvan} and the consistency of the $S^7$ Pauli reduction
was demonstrated in \cite{dewnic}.  The consistency of the $S^5$ Pauli 
reduction of type IIB supergravity has never been demonstrated, although
various non-trivial sub-cases have been explicitly shown to be consistent
(See, for example, \cite{cvluposatr}.)

   The three Pauli reductions mentioned above are all of considerable
complexity.  However, it turns out that we can consider further
(consistent) truncations within these examples, where only a small number 
of lower-dimensional fields are retained, which are nevertheless still
of a sufficiently non-trivial nature to be able to establish the results
that we wish to demonstrate.  Specifically, we shall first consider a 
truncation of the $S^7$ reduction of eleven-dimensional supergravity in 
which only the four-dimensional metric and a single scalar field are
retained.  This will enable us to provide an explicit example
in which it is impossible to give a reduction ansatz that is expressed in
terms of the fundamental fields of the higher-dimensional action, and so
it is not possible even in principle to discuss whether the 
lower-dimensional theory obtained by substituting the ansatz into the 
action agrees with that obtained by substituting instead into the 
equations of motion.

   As a second example, we shall elaborate the first case a little,
by including also a pseudo-scalar field in the truncated Pauli reduction.
Finally, for a third example, we shall consider a consistent truncation
of the $S^4$ Pauli reduction of eleven-dimensional supergravity.  In
this truncation, the only seven-dimensional fields that are retained are
the metric, a scalar, and a 3-form potential that satisfies a first-order
``odd-dimensional self-duality'' equation in $D=7$.  In this case, we shall
find that it {\it is} possible to express the ansatz in terms of the 
fundamental fields of an eleven-dimensional action, but nevertheless
one gets the incorrect seven-dimensional equations of motion
if one substitutes this ansatz into the action.  By contrast, in this as
in all the examples of consistent Pauli reductions, one does obtain the
correct lower-dimensional equations of motion by substituting the
ansatz into the higher-dimensional equations of motion.   

\section{Consistent Reduction on $S^7$}

\subsection{Truncation to gravity, scalar and pseudoscalar}

   The most complete presentation of the ansatz for the reduction of
the field equations of eleven-dimensional supergravity on $S^7$, 
yielding the field equations of maximal gauged
$SO(8)$ supergravity in four dimensions, was given in \cite{dewnic}.  
This contained an essentially complete proof of the consistency of the
reduction, the complete ansatz for the reduction of the fermions and
the metric reduction, and part of the reduction ansatz for the 4-form field 
strength.  Even had the 4-form ansatz been complete, the complexity
of the reduction procedure would make it very difficult to proceed with
an explicit discussion of substitution into the eleven-dimensional action.  
A very much simpler situation was considered in \cite{d4gauge}, where the
consistent reduction on $S^7$ that yields the truncation to $N=4$ gauged
$SO(4)$ supergravity was considered.  A complete and fully explicit ansatz
for the reduction of the metric and 4-form of eleven-dimensional supergravity
was obtained, and it was demonstrated that it consistently gave the required 
four-dimensional equations of motion when substituted into the 
eleven-dimensional equations of motion.  These equations follow from the
eleven-dimensional Lagrangian
\be
{\cal L}_{11} = {\hat R}\, {*\oneone} - \ft12 {\hat *\hat F_\4}\wedge \hat F_4
  + \ft16 \hat F_\4\wedge \hat F_\4\wedge \hat A_\3\,,\label{d11lag}
\ee
where $\hat F_\4\equiv d\hat A_\3$.  The bosonic fields in the
four-dimensional gauged $SO(4)$ supergravity comprise the metric, a 
scalar and a pseudoscalar, and the $SO(4)$ Yang-Mills fields.

    In order to simplify the discussion still further, while still
retaining a highly non-trivial reduction scheme, we shall begin by
setting the $SO(4)$ gauge fields to zero.  It is, of course, essential
that this truncation is itself a consistent one.  This follows from
a simple group-theoretic argument; since the metric, scalar and pseudoscalar
are $SO(4)$ singlets, it is necessarily consistent to set the $SO(4)$
gauge fields to zero.  The reduction ansatz for the remaining fields then
becomes (see \cite{d4gauge})
\bea
d\hat s_{11}^2 &=& \Delta^{2/3} ds_4^2 + 4g^{-2} \Delta^{2/3} \Big( d\xi^2 +
  \fft{c^2}{\Omega}\, d\Sigma_3^2 +
        \fft{s^2}{\wtd\Omega}\, d\wtd\Sigma_3^2\Big)\,,\label{metans}\\
\hat F_\4 &=& - g U \ep_\4 - 2sc g^{-1}\, {*d\phi}\wedge d\xi
+ 2sc g^{-1}\, \chi X^4 {*d\chi}\wedge d\xi  +\hat F_\4'\,,
\label{f4ans}
\eea
where $\ep_\4$ is the volume form of the four-dimensional metric $ds_4^2$,
\bea
X&=& e^{\ft12 \phi}\,,\qquad \wtd X= q\, X^{-1}\,,
                   \qquad q^2= 1+ \chi^2 X^4\,,\nn\\
\Omega &=& c^2 X^2 + s^2\,,\qquad \wtd\Omega= s^2 \wtd X^2 + c^2\,,
\qquad \Delta^2= \Omega\wtd\Omega\,,\qquad
  U= c^2 X^2 + s^2 \wtd X^2 + 2\,,\nn\\
c&=& \cos\xi\,,\qquad s=\sin\xi\,,
\eea
and $d\Sigma_3^2$ and $d\wtd\Sigma_3^2$ are metrics on two unit 3-spheres.
The extra term $\hat F_\4'$ in the expression for $\hat F_\4$ is given
by
\be
\hat F_\4' = d\hat A_\3'\,,\qquad \hat A_\3' = f\ep_\3 + \td f\td\ep_\3\,,
\ee
where $\ep_\3$ and $\td\ep_\3$ are the volume forms for the two 3-sphere
metrics, and
\be
f= \fft{c^4 \chi X^2}{g^3\Omega}\,,\qquad
\td f= -\fft{s^4\chi X^2}{g^3 \wtd\Omega}\,.
\ee
It is worth remarking that the metric on the unit round 7-sphere is given by
\be
d\Sigma_7^2 = d\xi^2 + c^2 d\Sigma_3^2 + s^2 d\wtd\Sigma_3^2\,,
\ee
which means that $\xi$ is a ``latitude coordinate,'' with 
$0\le\xi\le \ft12\pi$.  The surfaces
of constant $\xi$ are $S^3\times S^3$ with one $S^3$ shrinking to zero size
at $\xi=0$, whilst the other $S^3$ shrinks to zero size at $\xi=\ft12\pi$.
The scalar $\phi$ and pseudoscalar $\chi$ thus parameterise very specific
deformations of the 7-sphere, which maintain the homogeneity of
the $S^3\times S^3$ surfaces, in the reduction ansatz (\ref{metans}).

   It is easy to see that the Bianchi identity $d\hat F_\4=0$ is not
satisfied identically; rather, it implies a specific combination of
the scalar and pseudoscalar equations of motion.  Substitution of the
ansatze (\ref{metans}) and (\ref{f4ans}) into the full set of $D=11$
equations of motion following from (\ref{d11lag}), namely
\bea
\hat R_{MN} &=& \ft1{12} (\hat F^2_{MN} - \ft1{12} \hat F_\4^2\, \hat g_{MN})\,,
\label{einstein}\\
d{\hat *\hat F_\4} &=& \ft12 \hat F_\4\wedge \hat F_\4\,,\label{f4eom}
\eea
and Bianchi identity
\be
d\hat F_\4=0\,,\label{f4bianchi}
\ee
leads consistently to a set of
four-dimensional equations of motion, which can themselves be derived from
the Lagrangian
\be
{\cal L}_4= R {*\oneone} - \ft12 {*d\phi}\wedge d\phi -
      \ft12 e^{2\phi}\, {*d\chi}\wedge d\chi - V {*\oneone}\,,
\label{4dlag}
\ee
where the scalar potential $V$ is given by
\be
V= - g^2 (4 + 2 \cosh \phi + \chi^2\, e^\phi)\,.
\ee

\subsection{Truncation to the gravity plus scalar subsystem}

   We can make a further consistent truncation of the system described above,
in which the pseudoscalar $\chi$ is set to zero.  The ansatze for the metric
and 4-form then reduce to
\bea
d\hat s_{11}^2 &=& \Delta^{2/3} ds_4^2 + 4g^{-2} \Delta^{2/3} d\xi^2 +
 4 g^{-2} \Delta^{-1/3} \Big( c^2 e^{-\ft12\phi}  d\Sigma_3^2 +
       s^2\e^{\ft12\phi} d\wtd\Sigma_3^2\Big)\,,\label{metans2}\\
\hat F_\4 &=& - g U \ep_\4 - 2sc g^{-1}\, {*d\phi}\wedge d\xi \,,
\label{f4ans2}
\eea
where now $\Delta$ and $U$ have become
\be
\Delta = c^2 X + s^2 X^{-1}\,,\qquad U= c^2 X^2 + s^2 X^{-2} + 2\,,
\ee
with again $X= e^{\phi/2}$.

   We see that the Bianchi identity (\ref{f4bianchi}) implies the
four-dimensional equation of motion for the scalar field $\phi$, namely
\be
d{*d\phi}= g^2 (e^\phi - e^{-\phi} )\ep_\4\,.\label{d4scalareq}
\ee

   Since the 4-form field strength $\hat F_\4$ in (\ref{f4ans2}) is not 
closed {\it identically}, but only modulo the use of the four-dimensional
scalar field equation (\ref{d4scalareq}), it is clearly impossible to
re-express the 4-form ansatz in terms of its 3-form potential, since one
would need to be able to write $\hat A_\3$ off-shell.  Thus we have 
exhibited, in this relatively simple truncation of the $S^7$ reduction
ansatz, the fact that it cannot be written in such a way that it can
be substituted into the eleven-dimensional supergravity action.

\subsection{Dualisation in a toy model}\label{gravscaltoy}

   It is nonetheless interesting to observe that this particular
highly-truncated reduction ansatz can in fact be reinterpreted as an ansatz 
for a ``toy''
eleven-dimensional theory with a 7-form field strength rather than 
a 4-form field strength, for which one can then re-express the reduction
at the level of a fundamental 6-form potential.  The toy theory, it must
be emphasised, is {\it not} a dual formulation of eleven-dimensional 
supergravity; it is well known that it is {\it impossible} to rewrite
eleven-dimensional supergravity in terms of a dual formulation that involves
only a 6-form potential with no 3-form potential.  The obstacle to any such
rewriting is the Chern-Simons term
$\hat F_\4\wedge \hat F_\4\wedge \hat A_\3$, for which there is no way 
to avoid having $\hat A_\3$ as a fundamental field.  

    The reason why we
can give a reformulation in terms of a toy dual theory in the present 
case of the truncation that retains just gravity and the scalar field
in four dimensions is that the Chern-Simons term in the 
field equation (\ref{f4eom}) for
$\hat F_\4$ vanishes for this ansatz, and so we are left with just
$d{\hat *\hat F_\4}=0$ from this equation of motion.  Thus in this
truncated example we could take the ``original'' eleven-dimensional
theory to be described by the simpler Lagrangian
\be
{\cal L}_{11} = \hat R\, {\hat *\oneone} -\ft12 {\hat *\hat F_\4}\wedge 
{\hat F_\4}\,.
\ee
From the expressions in
\cite{d4gauge}, the dual of $\hat F_\4$ in our present truncation is
given by
\be
{\hat * \hat F_\4} = \fft{2 s^3 c^3 U}{g^6 \Delta^2 U}\,
d\xi \wedge \ep_\3\wedge \td\ep_\3 - \fft{s^4 c^4}{g^6 \Delta^2}\,
                 d\phi\wedge\ep_\3\wedge \td\ep_\3\,.
\ee
A simple calculation shows that the exterior derivative of ${\hat *F_\4}$ 
vanishes identically in this
truncated situation.   This means that it is natural to introduce the
dual 7-form field strength $\hat F_\7 = {\hat *\hat F_\4}$ here,
since then we can integrate the associated ansatz for $\hat F_\7$.
We find that $\hat F_\7 = d\hat A_\6$ where we can write
\be
\hat A_6 = \fft{s^4}{g^6}\, \Big( \fft{c^2 X^{-1}}{\Delta} +\fft12\Big)
 \ep_\3\wedge \td\ep_\3\,.\label{a6ans}
\ee

   In this simplified situation, therefore, we can consider a dualised theory 
where the 4-form field strength is replaced by a 7-form field strength, and
write the eleven-dimensional Lagrangian
\be
{\cal L}_{11} = \hat R {\hat *\oneone} - \ft12 {\hat *\hat F_\7}\wedge
              \hat F_\7\,,\label{d11duallag}
\ee
where ${\hat F_\7} \equiv d{\hat A_\6}$.  This may also be written
in terms of the Lagrangian density
\be
{\cal L}_{11} = \sqrt{-\hat g}\, \Big( \hat R - \fft1{2\cdot 7!}\,
         \hat F_\7^2\Big)\,.\label{f7lag}
\ee
Since we have an explicit ansatz for the fundamental 6-form potential
$\hat A_\6$, we may now investigate what happens if we substitute the
ansatz into the dualised 11-dimensional action following from 
(\ref{d11duallag}) or (\ref{f7lag}).  

   Substituting the ansatze (\ref{metans2}) and (\ref{a6ans}) into
(\ref{f7lag}), we obtain results as follows.  Firstly, we find that
\bea
\sqrt{-\hat g}\hat R &=& 2 s^3 c^3 g^{-7} Y  \sqrt{-g} R -
     \ft23 s^3 c^3 g^{-7} Y \sqrt{-g}\Big( 1 - \fft{2 s^2}{X\Delta}\Big)
    \square\phi \label{riccired}\\
&&- s^3 c^3 g^{-7} Y\sqrt{-g}\Big(
   1  + \fft{s^2}{3 X\Delta} - \fft{s^4}{3 X^2\Delta^2} \Big) (\del\phi)^2\nn\\
&&
   + s c g^{-5} Y\sqrt{-g}\Big(3 s^2 c^2(X^2+ X^{-2}) + \ft23 s^2(1+21 c^2)
   -\fft{s^2(3-4c^2)}{3 X\Delta} + \fft{s^4}{3 X^2\Delta^2}\Big)\,,\nn
\eea
where $Y$ denotes the product of the square roots of the determinants of the
two 3-sphere metrics $d\Sigma_3^2$ and $d\wtd\Sigma_3^2$.

   From the field strength, we get
\be
-\fft{1}{2\cdot 7!} \sqrt{-\hat g}\, \hat F_\7^2 = -s^5 c^5 g^{-7} Y\sqrt{-g}
 (\del\phi)^2 - s^3 c^3 Y \sqrt{-g}\fft{U^2}{g^5 \Delta^2}\,.
\ee

  It should be noted that unlike in a standard toroidal reduction,
we cannot drop the $\square\phi$ term in (\ref{riccired}) as a total
derivative, because there is $\phi$ dependence in the prefactor.  We can,
however, perform an integration by parts in the action, and this improves
the appearance of the scalar kinetic terms considerably.  However, we wish
to keep track of the $\xi$ dependence of the integrand in the action, and
so for now we shall integrate only over the four-dimensional spacetime.
Thus we find
\be
\int d^4 x {\cal L}_{11} = 2 s^3 c^3 g^{-7}\, Y\,
        \int d^4x \sqrt{-g}( R - \ft12 (\del\phi)^2 ) +
  g^{-5}\, Y\, \int d^4x \sqrt{-g}\, Q\,,\label{Qlag}
\ee
where
\be
Q= -\ft43 s^3 c(1-9c^2) + 2 s^3 c^3(X^2 + X^{-2}) -
  \fft{8 s^3 c^3}{3 X\Delta} + \fft{4 s^5 c}{3X^2\Delta^2}\,.\label{Qdef}
\ee

   At this stage, we see that although the integrand of the terms in
the action involving the lower-dimensional Einstein-Hilbert and scalar
kinetic terms has a uniform $\xi$ dependence, namely a prefactor
$s^3 c^3$, the same is not true of the terms involving the scalar
potential.  However, although this part of the integrand, given in
(\ref{Qdef}), has an extremely complicated $\xi$ dependence, it turns
out that after integrating over $\xi$ all the terms in the action assemble
into the hoped-for four-dimensional action, namely
\be
\int d^{11} x{\cal L}_{11} = \fft{128\pi^4}{3 g^7}\,
\int d^4x \sqrt{-g} \Big[ R - \ft12(\del\phi)^2 +
       g^2 (4 + e^\phi + e^{-\phi})\Big]\,.
\ee
This can be seen to be equivalent to the four-dimensional Lagrangian
(\ref{4dlag}), after truncating out the pseudoscalar $\chi$, which was itself
obtained by requiring that it reproduce the consistently-derived equations
of motion.

   The upshot of this discussion is that in the consistent
truncation where only the four-dimensional metric and scalar are
retained, one can in fact substitute the eleven-dimensional reduction
ansatz into the ``toy'' dualised eleven-dimensional action 
given by (\ref{d11duallag}) or
(\ref{f7lag}), and obtain the correct four-dimensional action. The way
in which this works is rather non-trivial, requiring cancels and
``conspiracies'' between terms in order to give the correct four-dimensional 
action.  It is also noteworthy that it is only after integrating over the
internal $S^7$ directions that one obtains a simple result; prior to
this integration, the eleven-dimensional integrand is not of the 
form of a single overall $S^7$-dependent function
multiplying a four-dimensional Lagrangian.

\subsection{The gravity plus scalar and pseudoscalar truncation}

   If we now go back to the somewhat larger reduction given in (\ref{metans})
and (\ref{f4ans}), where the pseudoscalar is included as well as the
scalar and the metric, we can see that neither the Bianchi identity nor
the field equation for the 4-form $\hat F_\4$ is identically satisfied.
(The Chern-Simons term in eleven dimensions does now give a contribution 
to the equation of motion for $\hat F_\4$.)   Each of the 
eleven-dimensional Bianchi identity and equation of motion 
implies that certain combinations of the four-dimensional equations
of motion must hold.  Thus one could not even construct a ``toy model''
for this larger truncation, analogous to the one in described in section 
\ref{gravscaltoy}, within which one could address the possibility of
substituting the ansatz into the action.  

\section{$S^4$ Reduction of $D=11$ Supergravity}

\subsection{Reduction of the field equations}

    The complete reduction of $D=11$ supergravity on $S^4$, yielding
${\cal N}=4$ $SO(5)$ gauged supergravity in $D=7$,  was obtained
in \cite{nasvamvan}.  The simpler case of the reduction to ${\cal N}=2$
$SU(2)$ gauged supergravity in $D=11$ was obtained in \cite{con1}.

   Here, we shall discuss a subset of the ${\cal N}=2$ reduction, in
which we consistently truncate out the $SU(2)$ Yang-Mills fields, leaving
just the metric, a dilaton and a 3-form gauge potential in $D=7$.

   After rescaling the gauge-coupling constant $g$ in \cite{con1} by sending
$g\rightarrow g/\sqrt2$ for convenience, the reduction ansatz for the
remaining fields can be written as
\bea
d\hat s_{11}^2 &=& \Delta^{1/3}\, ds_7^2 + 4 g^{-2}\,
        X^3 \Delta^{1/3}\, d\xi^2 + g^{-2}\, \Delta^{-2/3}\, X^{-1}\,
   c^2\, \sigma_i^2\,,\label{metans7}\\
\hat A_\3 &=&s\, A_\3 + f\, \ep_\3\,,\label{A3ans}
\eea
where $\sigma_i$ are left-invariant 1-forms of $SU(2)$,
$X=\exp(-\phi/\sqrt{10})$ where $\phi$ is the dilaton field, and
$\ep_\3=\sigma_1\wedge\sigma_2\wedge\sigma_3$.  The functions $\Delta$ and
$f$ are given by
\bea
\Delta &=& X^{-4} s^2 + X c^2\,,\\
f &=& g^{-3} (2 s + s c^2 \Delta^{-1}\, X^{-4}) \,,
\eea
and the symbols $s$ and $c$ are shorthand for
\be
s= \sin\xi\,,\qquad c=\cos\xi\,.
\ee
In \cite{con1}, the ansatz for the antisymmetric tensor was supplemented by
the requirement that the seven-dimensional 3-form satisfy the first-order
condition
\be
F_\4= \ft12 g X^{-4}\, {*A_\3}\,.\label{foans}
\ee
This could be viewed as part of the specification of the consistent 
reduction ansatz.  An 
alternative approach, as we shall discuss below, involves making a
modification to the
antisymmetric tensor ansatz so that it is expressed directly on the field
strength $\hat F_\4$ but can no longer be given for the potential $\hat A_\3$.  
One or other of these approaches is inevitable, because the 3-form
in seven dimensions should satisfy a first-order ``odd-dimensional 
self-duality'' equation, and it is not possible to derive a first-order
equation for $A_\3$ by substituting an ansatz for $\hat A_\3$ into
the second-order eleven-dimensional equation of motion for $\hat A_\3$.

   First, we note that the required $D=7$ equations of motion for the 
truncated gauged supergravity are
\bea
X^4\, {*F_\4} &=& - \ft12 g \, A_\3\,,\label{A3eq}\\
d(X^{-1}\, {*dX}) &=& \ft15 X^4\, {*F_\4}\wedge F_\4 - \ft1{10} g^2
    (X^{-8} + 2X^2 - 3 X^{-3})\, \ep_\7\,,\label{scaleq}\\
R_{\mu\nu} &=& 5 X^{-2} \del_\mu X\, \del_\nu X +
      \ft1{12} (F^2_{\mu\nu} - \ft3{20} F_\4^2\, g_{\mu\nu}\,,\label{meteq}
\eea
where $F_\4=dA_\3$.  Note that (\ref{A3eq}) is the ``first-order self-duality''
equation for the 3-form field.  By substituting (\ref{metans7}) and 
(\ref{A3ans}) into (\ref{einstein}) and (\ref{f4eom}), we derive the
scalar field equation (\ref{scaleq}) and the Einstein equation 
(\ref{meteq}).  

    As mentioned above, the first-order equation 
(\ref{A3eq}) is not itself {\it derivable} purely by using the ansatz
(\ref{A3ans}).  To see this, we note that (\ref{A3ans}) and (\ref{metans7})
imply
\bea
\hat F_\4 &=& s F_\4 + c d\xi\wedge A_\3 +
        \fft{\del f}{\del \xi}\, d\xi\wedge \ep_\3 +
      \fft{\del f}{\del X}\, dX\wedge \ep_\3\,,\nn\\
&=&  s F_\4 + c d\xi\wedge A_\3 
  -5 g^{-3} s c^4 \Delta^{-2} X^{-4} dX\wedge \ep_\3\nn\\
&&+ g^{-3} c^3 \Delta^{-2} X^{-8} [
 X^5(1+2X^5) c^2 + (4X^5-1)s^2] d\xi\wedge \ep_\3\,,\nn\\
{\hat *\hat F_\4} &=& 2 g^{-4} s c^3 \Delta^{-1}\,
                {* F_\4}\wedge d\xi\wedge \ep_\3  -
   \ft12 g^{-2}\, c^4 \Delta^{-1}\, X^{-3}\, {*A_\3}\wedge \ep_\3 \nn\\
 &&+\ft12 g^4\, c^{-3}\, \Delta^2 \fft{\del f}{\del \xi}\, \ep_\7
  - 2 g^2 \Delta^2 X^3 \, c^{-3}\,\fft{\del f}{\del X}\, {*dX}\wedge d\xi\,.
\label{f4starf4}
\eea
Substituting into the equation of motion for $\hat F_\4$ 
(\ref{f4eom}), we obtain the seven-dimensional equations:
\bea
d(X^{-4}{*A_\3}) &=& 0\,,\nn\\
dX\wedge (F_\4 - \ft12 g X^{-4}\,{* A_\3}) &=&0\,,\nn\\
5 dX\wedge ({*F_\4} + \ft12 g X^{-4}\, A_3) + gX^{-3}\, (1-X^5)
    (F_\4 - \ft12 g X^{-4}\,{* A_\3}) &=& 0\,,\nn\\
2 d(X^4 {*F_\4}) + 2g^2 X {*A_\3} - g( 4X^5-1) F_\4 &=&0\,,\nn\\
d(X^{-1} {*dX}) + \ft1{10} g^2 (X^{-8} + 2 X^2 - 3X^{-3}) \ep_\7 + \ft1{10}
   g F_4\wedge A_\3&=&0\,.\label{d7form}
\eea
(These equations arise from equating all the independent $\xi$-dependent
structures to zero.)   The first four equations in (\ref{d7form}) are
consistent with the first-order equation (\ref{A3eq}) (note that the
dual of the first-order equation gives $X^4 F_\4 = +\ft12 g {*A_\3}$).
However, the first four equations in (\ref{d7form}) do not actually
{\it imply} the first-order equation (\ref{A3eq}).  Finally, we can see that
the last equation in (\ref{d7form}) becomes, after the use of the first-order
equation, equivalent to the scalar equation (\ref{scaleq}).   For this
reason, the ansatz in \cite{con1} was supplemented by the condition
(\ref{foans}).  This was sufficient to provide a consistent embedding
of solutions of the seven-dimensional supergravity equations in eleven
dimensions. 

    We can in fact write a reduction ansatz that genuinely allows one to
{\it derive} the seven-dimensional equations of motion, including the
first-order equation for $A_\3$.   To achieve this, we must instead
write an ansatz for $\hat F_\4$ that {\it cannot} be written identically
as $d\hat A_\3$, \ie an ansatz that does not satisfy the Bianchi identity
$d\hat F_\4=0$ identically.   We do this by replacing $\hat F_\4$ in
(\ref{f4starf4}) by
\bea
\hat F_\4&=&  \ft12 g s X^{-4}\, {*A_\3} 
   + c d\xi\wedge A_\3 
 -5 g^{-3} s c^4 \Delta^{-2} X^{-4} dX\wedge \ep_\3\nn\\
&&
+ g^{-3} c^3 \Delta^{-2} X^{-8} [
 X^5(1+2X^5) c^2 + (4X^5-1)s^2] d\xi\wedge \ep_\3\,.\label{F4ans}
\eea
(Note that what we have done here is to use the first-order
equation (\ref{A3eq}) to replace $F_\4$ by $\ft12g X^{-4}\, {*A_\3}$ in
the first term.)  It is easy to see that the Bianchi identity 
$d\hat F_\4=0$ indeed now implies (as well as previously-obtained equations)
the first-order equation (\ref{A3eq}).  In all other respects, the modified
ansatz yields the same conclusions as previously, and so we now have
a reduction scheme from which one can consistently {\it derive} the
lower-dimensional equations of motion.  It should be emphasised, though,
that this is a reduction in which the ansatz is now given for $\hat F_\4$,
but it cannot be given for $\hat A_\3$.

   Clearly this modified reduction ansatz cannot be substituted into the
usual eleven-dimensional action given by (\ref{d11lag}), precisely 
because it can no longer be given as an ansatz for the 3-form potential
$\hat A_\3$.  One could, of course, substitute the ansatz in its original
form (\ref{metans7}) and (\ref{A3ans}) into the action, but since this
ansatz does not even give a {\it derivation} of the first-order equation
for $A_\3$ when substituted into the eleven-dimensional field equations,
one can hardly expect it to work when substituted into the action. Indeed,
an explicit calculation shows that
one does not obtain a sensible seven-dimensional action upon integrating
over the 4-sphere.

\subsection{Substitution into the eleven-dimensional action}

   A new first-order formulation of eleven-dimensional supergravity 
was constructed in \cite{nasvamvan}. This entails writing the bosonic 
Lagrangian as
\be
{\cal L} = {\hat R}{\hat *\oneone}  + \ft12 {\hat *\hat{\cal F}_\4} \wedge
\hat{\cal F}_\4 - {\hat *\hat{\cal F}_\4}\wedge d\hat A_\3 +
 \ft16 d\hat A_\3\wedge d\hat A_\3 \wedge \hat A_\3\,.\label{folag}
\ee
where $\hat{\cal F}_\4$ and $\hat A_\3$ are treated as independent fields.
The equation of motion for $\hat{\cal F}_\4$ implies
\be
\hat{\cal F}_\4 = d\hat A_\3\,,\label{foeq1}
\ee
while the equation of motion for $\hat A_\3$ implies
\be
d{\hat *{\cal F}_\4} = \ft12 d\hat A_\3\wedge d\hat A_\3\,.\label{foeq2}
\ee

    We could now take the ansatz for $\hat A_\3$ to be given by (\ref{A3ans}),
and the ansatz for $\hat{\cal F}_\4$ to be given by (\ref{F4ans}), \ie
\bea
\hat {\cal F}_\4&=&  \ft12 g s X^{-4}\, {*A_\3} + c d\xi\wedge A_\3 
 -5 g^{-3} s c^4 \Delta^{-2} X^{-4} dX\wedge \ep_\3\nn\\
&&
+g^{-3} c^3 \Delta^{-2} X^{-8} [
 X^5(1+2X^5) c^2 + (4X^5-1)s^2] d\xi\wedge \ep_\3\,.\label{cF4ans}
\eea
It is certainly the case that if one substitutes these ans\"atze into
the equations of motion (\ref{foeq1}) and (\ref{foeq2}) following from
(\ref{folag}), then one derives the correct seven-dimensional
equations of motion (including the first-order equation (\ref{A3eq})).

   However, if we substitute the ans\"atze (\ref{metans7}), 
(\ref{A3ans}) and (\ref{cF4ans}) into
the first-order Lagrangian (\ref{folag}), we find (focusing for now on
just the terms involving $A_\3$) the terms
\bea
{\cal L}&=& -\ft14 g^{-2}\, c^3 \, X^{-4}\, {*A_\3}\wedge A_\3\wedge d\xi\wedge
   \ep_\3 \nn\\
&&+ g^{-3}\, (s^2 c^3 \Delta^{-1}\, X^{-4} -
      \ft23 s^4 c \Delta^{-2}\, X^{-8} )\, F_\4\wedge A_\3\wedge d\xi \wedge
     \ep_\3 + \cdots\,,\label{lag4}
\eea
where the ellipses indicate terms involving fields other than $A_\3$. 
If the integration over $\xi$ (and, trivially, over the $S^3$) were to
yield appropriate quantities, this
would have the possibility to yield a seven-dimensional Lagrangian with
terms of the desired form
\be
{\cal L}_7 \sim -\ft12 X^{-4}\, {*A_\3}\wedge A_\3 + g^{-1}\, F_\4\wedge A_\3
   + \cdots\,.
\ee
The equation of motion from this Lagrangian would produce the desired
first-order equation (\ref{A3eq}).  What actually happens, however, is
that the integration over the prefactor of $F_\4\wedge A_\3$ in
(\ref{lag4}) yields precisely 0.  In fact, therefore, one obtains only
\be
{\cal L}_7 = -\ft12 X^{-4}\, {* A_\3}\wedge A_\3 + \cdots\,.
\ee
Thus we obtain a seven-dimensional Lagrangian that does {\it not} produce
the proper seven-dimensional equations of motion.  This is a clear-cut
example where we have an explicit ansatz for the fundamental fields in
a higher-dimensional action, which nevertheless fails to give the
correct lower-dimensional action.

\section{Discussion and Conclusions}

   A consistent Pauli reduction is defined to be a dimensional reduction
on a coset space $G/H$ which retains a finite number of lower-dimensional
fields including the gauge bosons of the isometry group $G$.  Substitution
of the reduction ansatz into the higher-dimensional equations of motion
yields a consistent system of equations of motion for the lower-dimensional
fields.  Conversely, any solution of the lower-dimensional equations of
motion will lift to give a solution of the original higher-dimensional
equations of motion.

   In this paper, we have used several relatively simple examples to
study the question of whether a consistent reduction ansatz of this type 
can alternatively be substituted into the higher-dimensional action, 
thereby yielding a lower-dimensional action that correctly reproduces the
lower-dimensional equations of motion obtained as described above.  This
question is non-trivial for a variety of reasons, not the least of which
is that there is no known group-theoretical explanation for why 
consistent Pauli reductions have to work; they are very much the exception
rather than the rule, and their success depends on very special features 
and ``conspiracies'' that arise in specific higher-dimensional theories. 
In the absence of a group-invariance argument of the type one has in a
more tradiational Kaluza-Klein circle reduction or DeWitt group-manifold
reduction, there is apparently no {\it a priori} reason why a substitution
of the Pauli reduction ansatz into the higher-dimensional action should
produce a correct lower-dimensional action.

   In order to render the computations manageable, our 
examples have been chosen to be further truncations of certain highly
non-trivial Pauli reductions, namely the $S^7$ and $S^4$ reductions of
eleven-dimensional supergtravity.   

   In some cases, there are clear-cut reasons why one cannot even begin
to discuss the substitution of the reduction ansatz into the action.  Our
first example, the truncation of the $S^7$ reduction to just the
four-dimensional metric and one scalar field, illustrates this rather
clearly.  The reduction ansatz for the antisymmetric tensor of 
eleven-dimensional supergravity can only be given for the field strength
$\hat F_\4$, and there is no way to write an off-shell expression for the 
fundamental gauge potential $\hat A_\3$.  Since there is no formulation
of an action for eleven-dimensional supergravity that does not include
$\hat A_\3$ as a fundamental field (because of the $\hat F_\4\wedge 
\hat F_\4\wedge \hat A_\3$ Chern-Simons term), one therefore has no 
possibility of
substituting an ansatz for this system into the action.  Of course, the 
conclusion applies all the more to the case where one retains the full set
of fields of the $S^7$ Pauli reduction (\ie the fields of four-dimensional 
maximal gauged supergravity).

   In our example discussed above, where only the four-dimensional metric
and a scalar field are retained, it so happens that the eleven-dimensional
Chern-Simons term plays no r\^ole in the equations of motion.  In this
case, therefore, one can consider a ``toy'' theory, namely the bosonic sector 
of eleven-dimensional supergravity with the Chern-Simons term omitted.  
It is furthermore the case for this gravity plus scalar truncation that 
although one cannot give a reduction ansatz for the potential $\hat A_\3$
(since the Bianchi identity for $\hat F_\4$ in the field-strength reduction
ansatz is not identically satisfied, but instead implies the lower-dimensional
scalar equation of motion), the higher-dimensional equation of motion
for $\hat F_\4$ {\it is} identically satisfied.  This means that within the
framework of the ``toy'' theory one can dualise and re-express it in terms
of a 6-form potential $\hat A_\6$ instead, and then one can give an
explicit reduction ansatz for the dualised gauge potential $\hat A_\6$
itself.  Having done this, one can investigate what happens if this is
substituted into the dualised ``toy'' action; it does in fact turn out
to yield (in a rather non-trivial way) the correct four-dimensional action.

    This trick of passing to a toy theory where a dualisation is possible
is very specific to the highly-truncated gravity plus scalar system.  We
next showed that if one considers instead a slightly less extreme 
truncation of the $S^7$ reduction ansatz, where a pseudoscalar is
retained as well as the metric and the scalar field, then neither 
the Bianchi identity nor the field equation for $\hat F_\4$ is identically
satisfied, the Chern-Simons term of eleven-dimensional supergravity
now plays an essential r\^ole, and there is no possibility of recasting the
ansatz in a manner that would allow it to be substituted into the
eleven-dimensional action.

   One might have drawn the conclusion from the examples above that at least
in those cases where
a Pauli reduction ansatz could be given for the fundamental fields 
appearing in the higher-dimensional action, then substitution of the
ansatz into the action would give the correct lower-dimensional action
(as in the case of the gravity plus scalar truncation in the toy theory
described above).  A third example that we studied appears to invalidate 
this conclusion.  We considered a truncation of the $S^4$ Pauli reduction
of eleven-dimensional supergravity, in which just the metric, a scalar
field, and a 3-index antisymmetric tensor $A_\3$ are retained.  The
3-form $A_\3$ satisfies a first-order ``odd-dimensional self-duality
equation'' in seven dimensions.  We showed that the consistent reduction
ansatz could be written explicitly in terms of the fundamental fields of
a recent first-order reformulation of eleven-dimensional supergravity
given in \cite{nasvamvan}.  Nonetheless, upon substitution into the 
eleven-dimensional action, this failed to produce the correct seven-dimensional
action.

   The moral to be drawn from these examples seems to be that there is 
much that remains to be understood about the remarkable examples where
consistent Pauli reductions occur.  The usual expectation, derived from
experience with Kaluza-Klein or DeWitt group-manifold reductions where
group-invariance arguments apply, apparently do not extend to the much
less well understood Pauli reductions.  This leaves many open questions
and interesting avenues for further investigation.

\section*{Acknowledgement}
 
    C.N.P. and K.S.S are grateful to the Theory Division at CERN for
hospitality during the course of this work.

\end{document}